\documentclass[pra,twocolumn,groupedaddress,amsmath,amssymb,aps,10pt,showpacs]{revtex4-1}
\DeclareMathOperator{\sgn}{sgn}
\usepackage[sort&compress]{natbib}
\usepackage{graphicx,color,bm}
\begin{document}

\title{Perturbative correction to the ground state properties of one-dimensional strongly interacting bosons in a harmonic trap}
\author{Francis N.~C.~Paraan}
\email[Electronic address: ]{\url{fparaan@ic.sunysb.edu}}
\author{Vladimir E.~Korepin}
\email[Electronic address: ]{\url{korepin@max2.physics.sunysb.edu}}
\affiliation{C. N. Yang Institute for Theoretical Physics, State University of New York at Stony Brook, NY 11794-3840, USA}
\date[Received ]{\today}

\begin{abstract}
We calculate the first-order perturbation correction to the ground state energy and chemical potential of a harmonically trapped boson gas with contact interactions about the infinite repulsion Tonks-Girardeau limit. With $c$ denoting the interaction strength, we find that for a large number of particles $N$ the $1/c$ correction to the ground state energy increases as $N^{5/2}$, in contrast to the unperturbed Tonks-Girardeau value that is proportional to $N^2$. We describe a thermodynamic scaling limit for the trapping frequency that yields an extensive ground state energy and reproduces the zero temperature thermodynamics obtained by a local density approximation.
\end{abstract}

\maketitle
\hbadness=10000

\section{Introduction}
The realization of quasi one-dimensional ultracold boson gases with tunable interaction parameters \cite{gorlitz,paredes,kinoshita} and the succeeding advances in atom chip trap technology \cite{horikoshi,dellapietra,vanamerongen,vanes} have renewed interest in theoretical models of one-dimensional bosons with short-ranged interactions. Of particular relevance to workers in this field is the Lieb-Liniger model \cite{lieb1,*lieb2} in which contact interactions are described by Dirac delta functions. The suitability of this model in describing the low temperature properties of these quasi one-dimensional bosonic systems has been further strengthened by Olshanii's analysis of the low energy scattering of atoms under tight transverse harmonic confinement: The longitudinal $s$-wave scattering amplitudes are indeed reproduced by a one-dimensional pseudopotential proportional to a Dirac delta function \cite{dunjko}. The resulting demonstration that the magnitude of this effective delta interaction can be explicitly calculated from the three-dimensional atomic scattering length and the dimensions of the confining external trap \cite{olshanii1} further strengthens the link between theoretical one-dimensional models and quasi one-dimensional experiments.

Still, the free Lieb-Liniger model is quite an idealization for actual experiments because the atoms are generally longitudinally confined by an external potential and thus much effort has been devoted to studying the effects of confinement of interacting bosons \cite{dunjko,yang1,*yang2,cherny}. Introducing an external harmonic potential to the free Lieb-Linger model of spinless bosons leads to the many-particle Schr\"odinger eigenvalue equation
\begin{multline}
\bar{E}^\text{b} \Psi^\text{b} =\Biggl[\sum_{i=1}^N-\frac{\hbar^2}{2m}\frac{\partial^2}{\partial x_i^2} + \frac{m\omega^2 x_i^2}{2}\Biggr]\Psi^\text{b} \\- \frac{\hbar^2}{ma}\sum_{i<j} \delta(x_j-x_i)\Psi^\text{b}\negmedspace,
\label{ham0}
\end{multline}
where $m$ is the mass of each of the $N$ atoms, $\omega$ is the angular frequency of the trap, and $a$ is the one-dimensional scattering length. The superscript `b' refers to the bosonic nature of the labeled quantities. Measuring energy in units of $\hbar\omega$ and length in oscillator units $\ell=\sqrt{\hbar/m\omega}$ gives the dimensionless eigenvalue equation
\begin{equation}
{E}^\text{b} \Psi^\text{b} = \Biggl[\sum_{i=1}^N-\frac{1}{2}\frac{\partial^2}{\partial x_i^2} + \frac{1}{2}x_i^2 + c\sum_{i<j} \delta(x_j-x_i)\Biggr]\Psi^\text{b},
\label{ham}
\end{equation}
where we have introduced the dimensionless interaction strength $c=-\ell/a$. We consider here the repulsive case $c>0$ (negative scattering length) to be specific. In the absence of an harmonic potential the corresponding eigenvalue equation is solvable by the Bethe ansatz and consequently much is known about the ground state and elementary excitations of this system \cite{lieb1,*lieb2}, as well as the properties of the various correlation functions at zero and finite temperatures \cite{korepin}. However, for the important case of harmonic confinement an exact solution to this problem for general values of the interaction strength $c$ is lacking. The sole exceptions are the two-particle case that is separable in relative and central coordinates \cite{busch}, and the Tonks-Girardeau (TG) limit of infinite repulsion $c\to +\infty$ in which the exact $N$-particle wavefunctions are absolute values of Slater determinants \cite{girardeau1,*girardeau2}. These results, especially the latter, are prototypical examples of the fermion-boson duality derived by Girardeau \cite{girardeau3} and later generalized by Cheon and Shigehara \cite{cheon} for one-dimensional systems of particles having contact interactions. For finite values of the interaction parameter $c$, expressions for the atomic density and collective oscillation frequencies have been calculated using local density approximations \cite{dunjko,kolomeisky,menotti,kim,yang1,*yang2} and time-dependent density functional theory \cite{kim,brand}, while formal expressions for the self-consistent Hartree-Fock equations for the single-particle density matrix have been obtained for general trapping potentials \cite{cherny}. The analogous problem of a system of confined interacting fermions has also been treated by similar methods \cite{granger,*astrakharchik,*tokatly,*magyar1,*xianlong,*yang4}.

In this work, we will use the mentioned fermion-boson relation to develop perturbative $1/c$ corrections to the ground state energy and chemical potential of an harmonically confined interacting boson gas about the Tonks-Girardeau solution. The details of the specific fermion-boson mapping we employ here that utilizes a non-local pseudopotential \cite{cherny} are given in Section \ref{fbmapping}.  In Section \ref{perturb} we present our perturbation calculations for the general case of an $N$-particle system obtaining a closed form analytical result that is calculable for any $N$. We analyze few-body cases and discuss the thermodynamic limit $N\to\infty$ of our solution in Section \ref{cases}. We summarize our results and give concluding remarks in Section \ref{conclusion}.

\section{Fermion-boson mapping}\label{fbmapping}
In one dimension it has been demonstrated that a bosonic model with pairwise contact interactions of strength $c$ can be mapped into a fermionic model with pairwise interactions of strength $1/c$ \cite{cheon}. Specifically, given a fermionic wavefunction $\Phi^\text{f}$ that satisfies the eigenvalue equation
\begin{equation}
\Biggl[\sum_{i=1}^N-\frac{1}{2}\frac{\partial^2}{\partial x_i^2} + \frac{1}{2}x_i^2 + \hat{V}^\text{f}\Biggr] \Phi^\text{f} = E^\text{f}\Phi^\text{f},
\label{hamf}
\end{equation}
an appropriate choice of a pseudopotential operator $\hat{V}^\text{f}$ allows us to make the following correspondence between the bosonic wavefunction $\Psi^\text{b}$ and its fermionic counterpart $\Phi^\text{f}$
\begin{equation}
\Psi^\text{b} = \mathcal{A}\Phi^\text{f}(x_1, \dotsc, x_N), 
\end{equation}
where $\mathcal{A} \equiv \prod_{i<j}\sgn(x_j-x_i)$ is a function that is completely antisymmetric under any transposition $x_i \leftrightarrow x_j$. Since the complex squares of the wavefunctions $\Psi^\text{b}$ and $\Phi^\text{f}$ are identical, the energy eigenvalues $E^\text{b}$ and $E^\text{f}$ are equal. For our specific problem, a suitable fermionic pseudopotential operator has matrix elements in the coordinate representation given by \cite{cherny}
\begin{equation}\label{vferm}
\langle \varphi^\text{f} |\hat{V}^\text{f}|\phi^\text{f}\rangle = -\frac{4}{c}\sum_{i<j}\int\lim_{r_{ij}\to 0}\biggl[\frac{\partial\varphi^\text{f\,*}}{\partial r_{ij}}\times\frac{\partial\phi^\text{f}}{\partial r_{ij}}\biggr]\, dR_{ij},
\end{equation}
where $r_{ij} = x_j-x_i$ and $R_{ij}=\tfrac{1}{2}(x_j+x_i)$ are relative and central coordinates, and $\varphi^\text{f}(x_1,\dotsc,x_N)$ and $\phi^\text{f}(x_1,\dotsc,x_N)$ are the coordinate space wavefunctions corresponding to the fermionic state kets $|\varphi^\text{f}\rangle$ and $|\phi^\text{f}\rangle$.

In the infinite repulsion limit, the bosonic eigenvalue equation \eqref{ham} is solved by the absolute value of the ground state Slater determinant \cite{girardeau1,*girardeau2}
\begin{equation}
\Psi_\text{TG}^\text{b} = \frac{1}{\sqrt{N!}} \bigl| \det \psi_n(x_m)\bigr| \equiv \mathcal{A}\Phi_\text{TG}^\text{f},
\label{wf1}
\end{equation}
where  $\Phi_\text{TG}^\text{f}$ is the fermionic ground state wavefunction, $\{x_m\}$ are the coordinates of the atoms and $\{\psi_n\}$ are the $N$ lowest energy single-particle harmonic oscillator eigenfunctions $\psi_n(x) = \pi^{-1/4}(2^nn!)^{-1/2}H_n(x)e^{-x^2/2}$. The $H_n(x)$ appearing here are the usual Hermite polynomials. The corresponding energy of this TG ground state (in units of $\hbar\omega$) is 
\begin{equation}
E_\text{TG}^\text{b} = \tfrac{1}{2}N^2.
\end{equation}
For a finite and large repulsion strength we may therefore use the quantity $1/c \ll 1$ as a perturbation parameter for the fermionic problem \eqref{hamf} so that ordinary first-order perturbation theory gives the desired correction
\begin{equation}
E_0^\text{b} = E_0^\text{f} = \tfrac{1}{2}N^2 + \langle\Phi_\text{TG}^\text{f}|\hat{V}^\text{f}|\Phi_\text{TG}^\text{f}\rangle + \mathcal{O}(1/c^2).
\end{equation}

\begin{figure}[tb]
	\centering
		\includegraphics[width=0.8\linewidth]{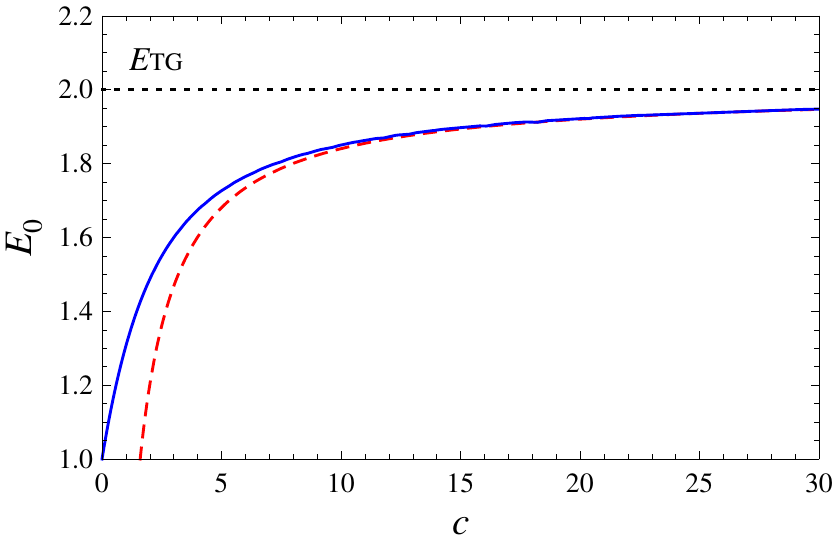}
	\caption{The first-order perturbation result (dashed line) for the ground state energy of two delta interacting bosons in a harmonic trap is compared to the exact solution (solid line). The ground state energy $E_\text{TG}$ in the infinite repulsion limit is given by the horizontal dotted line.}\label{2body}
\end{figure}

\section{Ground state energy correction}\label{perturb}
In this section we work mainly in the fermionic sector and omit the superscripts `f' for brevity. Here, our objective is to explicitly evaluate the leading correction $\delta E \equiv\langle\Phi_\text{TG}|\hat{V}|\Phi_\text{TG}\rangle$ to the ground state energy $E_0$. Since $\Phi_\text{TG}$ is a Slater determinant and the fermionic interaction operator $\hat{V}$ is a sum of two-body operators $\hat{v}$ having matrix elements
\begin{multline}\label{vklmn}
{v}_{klmn} = -\frac{4}{c}\int\lim_{r\to 0}\Biggl\{\frac{\partial\bigl[\psi_k(x_1)\psi_l(x_2)\bigr]^*}{\partial r}\\ 
\times \frac{\partial\bigl[\psi_m(x_1)\psi_n(x_2)\bigr]}{\partial r}\Biggr\}\, dR,
\end{multline}
we may calculate the perturbation $\delta E$ using the Slater-Condon rule $\delta E = \sum_{k<l} ( v_{klkl}-v_{kllk})$ \cite{slater,*condon}. Prior to calculating the derivatives appearing inside the integral \eqref{vklmn}, we must be careful to write the coordinates $x_1 = R + \tfrac{1}{2}r$ and $x_2 = R - \tfrac{1}{2}r$ in terms of the relative and central coordinates $r$ and $R$. The symmetry of the integrand allows us to write and define $v_{klkl} = -v_{kllk} \equiv \tilde{v}_{kl}$ where
\begin{equation}
\tilde{v}_{kl} = -\frac{4}{c}\int\lim_{r\to 0}\Biggl\{\frac{\partial\bigl[\psi_k(x_1)\psi_l(x_2)\bigr]}{\partial r}\Biggr\}^2\, dR.
\end{equation}
Thus, the leading correction becomes $\delta E =2\sum_{k<l}\tilde v_{kl}$ and is always negative as expected. After some manipulation, we obtain a finite series expression for the energy correction that may be evaluated for any number of atoms $N$:
\begin{align}\label{corrn}
\delta E &= \frac{1}{c}\sqrt{\frac{2}{\pi^3}}\sum_{l=1}^{N-1}\frac{\Gamma\bigl(l-\tfrac{1}{2}\bigr)}{\Gamma(l+1)}\nonumber \\
&\qquad\times\sum_{k=0}^{l-1}\frac{(l-k)^2\Gamma\bigl(k-\tfrac{1}{2}\bigr)}{\Gamma(k+1)} \, _3F_2\biggl[\begin{subarray}{l}
\tfrac{3}{2},\, -k,\, -l\\ \tfrac{3}{2}-k,\, \tfrac{3}{2}-l
\end{subarray};1\biggr]. 
\end{align}

\begin{figure*}[tb]
	\centering
		(a)\includegraphics[width=0.45\linewidth]{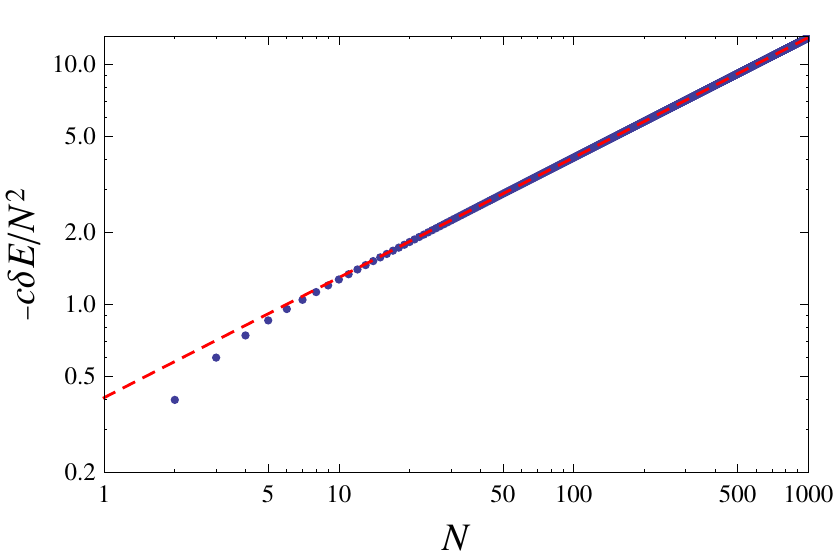}(b)\includegraphics[width=0.45\linewidth]{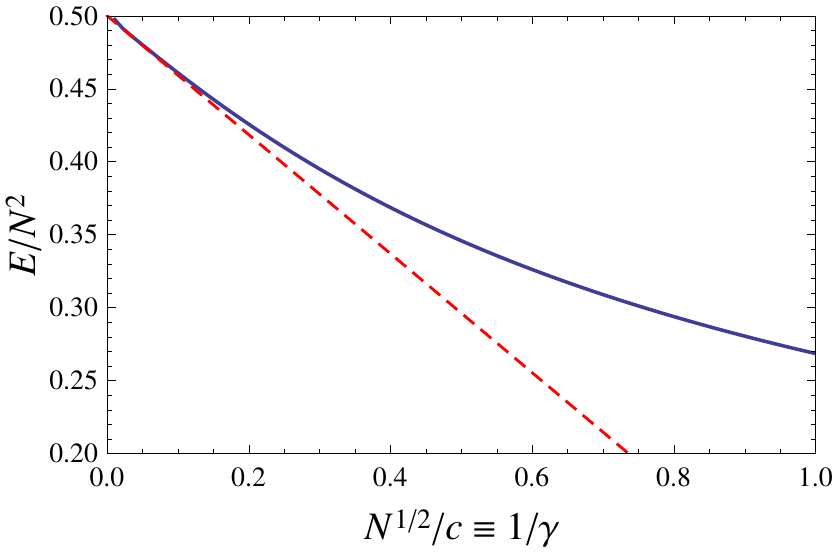}
	\caption{(a) The scaled perturbation $-c\delta E/N^2$ (dots) grows as a power law $\sqrt{N}$ in the limit $N\to\infty$. (b) Our large $N$ result is consistent with numerical calculations (solid line) in the Thomas-Fermi approximation for $\sqrt{N}/c\ll 1$. The dashed lines are regression fits $c\delta E \approx -0.408N^{5/2}$ calculated from values $N\in[100,1000]$. The prefactor $-0.408$ is independent of the particular values chosen in this interval.}\label{energyn}
\end{figure*}

\section{Few particle results and thermodynamic limit}\label{cases}
The special case of $N=2$ particles is separable in relative and central coordinates and the resulting eigenvalue problem for the ground state energy is exactly solvable. Upon imposing vanishing boundary conditions on the two-body wavefunction at infinity, we find that the ground state energy $E_0$ of the trapped two boson system satisfies the transcendental equation
\begin{equation}
\frac{2\,\Gamma\bigl[\tfrac{1}{2}(1+E_{0})\bigr] \tan\bigl[\tfrac{1}{2}(1-E_{0})\pi\bigr]}{\Gamma\bigl[E_{0}/2\bigr]} = -\frac{c}{\sqrt{2}}, 
\end{equation}
where $E_{0}\in[1,2]$. Our perturbative result \eqref{corrn} gives $\delta E = -(2/c)\sqrt{2/\pi}$ and coincides with the leading correction term in the $1/c$ series expansion of this exact solution about the TG ground state energy $E_\text{TG} = 2$. Also, an essentially identical result was obtained by Sen~\cite{sen} for the reduced single-particle problem using a different local pseudopotential involving the second derivative of the delta function $\propto \delta''(x_2-x_1)$ \footnote{As noted in Ref.~\cite{cherny}, Sen's potential and the pseudopotential used here yield identical interaction matrix elements between states described by continuous fermionic wavefunctions. Since our calculations involve continuous Slater determinants, the distinction between pseudopotentials is not relevant here.}. A comparison of our first-order perturbation result and the exact two-particle ground state energy is shown in Figure \ref{2body} and we observe good agreement between the two results in the strongly interacting regime $c\gg10$. 

Before discussing the situation for more than two particles, let us restore units and rewrite the perturbed ground state energy as
\begin{equation}\label{ezeron}
E_0(N) = \tfrac{1}{2}\hbar\omega N^2\bigl[1 + {2\alpha(N)}/{c}\bigr] + \mathcal{O}(1/c^2),
\end{equation}
where $\alpha(N)$ is a dimensionless function of $N$. For values of $N$ up to $10^3$ the magnitude of the scaled first-order correction $-c\delta E(N)/N^2$ is plotted in Figure \ref{energyn}a as a function of $N$ on a double logarithmic plot. Inspection of this graph suggests a simple power law scaling for the first-order correction with large $N$. Leading order asymptotic analysis reveals that the partial sums in Eq.~\eqref{corrn} scale as $\sim\negthinspace N^{3/2}$ for large $N$ so that $N^2\alpha(N) \sim N^{5/2}$:
\begin{equation}\label{ezeroasympt}
E_0(N) \approx \tfrac{1}{2}\hbar\omega N^2\bigl[1 + 2\alpha_0\sqrt{N}/{c}\bigr],\quad N \to\infty,
\end{equation}
where $\alpha_0$ is a constant number. Indeed, for as few as $N\gtrsim15$ particles the factor $\alpha(N)$ is quite well approximated (within 1\%) by the function $\alpha_0\sqrt{N}$ with $\alpha_0 \approx -0.408$. In other words, the correction factor $\alpha(N)$ reaches its asymptotic scaling behavior for systems as small as $N\gg\mathcal{O}(10^1)$. 

To obtain a thermodynamic limit with an extensive ground state energy, we observe that in addition to sending the number of particles to infinity we must also require the trapping frequency $\omega$ to vanish as $1/N$. As we shall see, this condition allows us to reproduce the Thomas-Fermi results near the Tonks-Girardeau limit. This scaling requirement implies that the quantity $\ell/\sqrt{N}$ approaches a constant in the thermodynamic limit, in contrast to the thermodynamic limit in the case of a Lieb-Liniger gas confined in a flat-bottomed box, in which the linear dimension of the system is taken to scale proportionally with particle number \cite{batchelor}. Looking back at our asymptotic expression $\alpha(N)\approx\alpha_0\sqrt{N}$, we find that the quantity $\sqrt{N}/c$ approaches a constant value as $N\to\infty$ in our prescribed thermodynamic limit. This is precisely the condition used by Ma and Yang \cite{yang1,*yang2} to obtain the ground state energy of the trapped interacting boson gas in the Thomas-Fermi formalism, which we reproduce here in Figure \ref{energyn}b. We find that our first-order $1/c$ result is reliable for $\sqrt{N}/c \lesssim 0.1$, which means that for a typical experimental setup with hundreds of atoms first-order perturbation theory and the Thomas-Fermi result coincide only in the extreme repulsion limit $c\gtrsim10^2$.  

If we now define the Tonks-Girardeau limit chemical potential as $\mu_\text{TG} \equiv \lim_{N\to\infty,\,\omega\to 0}\hbar\omega N$ and the scaled interaction parameter as $\gamma \equiv  \lim_{N\to\infty,\,\omega\to 0}	c/\sqrt{N}$, we obtain the zero temperature chemical potential 
\begin{equation}
\mu \approx \mu_\text{TG}\bigl[1 + \tfrac{5}{2}\alpha_0/\gamma\bigr].
\end{equation}
The first term in this expression corresponds to the chemical potential of free fermions in a one-dimensional harmonic trap while the last term gives the reduction in the chemical potential due to the finite repulsion correction and serves as a measure of the departure of the system from the unitarity limit.

\section{Concluding remarks}\label{conclusion}
In this work we have calculated the first-order finite repulsion correction to the ground state energy of harmonically trapped bosons having contact interactions for any finite number of particles $N$. For $N\gg\mathcal{O}(10^1)$ we found that for a fixed interaction strength this correction scales as a power law $N^{5/2}$, which allowed us to describe a thermodynamic limit that reproduces known results from Thomas-Fermi approaches. This contribution clarifies the smooth transition of the ground state properties of an harmonically confined interacting boson system as the particle number goes from finite $N$ to infinity near the Tonks-Girardeau limit. We have demonstrated that in this strongly interacting regime, to at least leading order in $1/c$, finite number effects are negligible in current experimental situtations that have $\sim\negthinspace 10^2$ atoms.

A natural extension of this work would involve higher order corrections to the ground state energy and many-body wavefunction, as was done recently for a wedge-shaped trapping potential \cite{jukic}. If we take the set of all fermionic Slater determinants as an expansion basis for ordinary perturbation theory about the TG limit, we discover that the perturbing pseudopotential couples the ground state to an infinite number of excited states. We therefore expect a complicated analytical result for the second order energy correction resulting in a numerical problem that may require a truncation of the corresponding Hilbert space. However, on the basis of the agreement between our asymptotic results and the Thomas-Fermi calculation (Figure \ref{energyn}b), we conjecture that the second order correction scales as $N^3/c^2 > 0$ in the thermodynamic limit.

\subsection*{Acknowledgments}
During the course of this work many insightful discussions were shared with A.~G.~Abanov, D.~Schneble, and F. Franchini. F.~P.~is financially supported by a Teaching Assistantship from the Department of Physics and Astronomy of the State University of New York at Stony Brook. V.~K.~acknowledges financial support by NSF Grant No. DMS-0905744.
\providecommand{\bibyu}{Yu} \providecommand{\bibth}{Th}

\end{document}